\documentclass[prd,showpacs,superscriptaddress,amsfonts,amssymb,amsmath,twocolumn]{revtex4-2}

\usepackage{epsfig}
\usepackage{graphics}
\usepackage{amsmath}
\usepackage{xcolor}
\usepackage{soul}

\date{\today}

\newcommand{\ud}{\mathrm{d}}
\newcommand{\ui}{\mathrm{i}}
\newcommand{\ue}{\mathrm{e}}

\begin{document}

%%%%%%%%%%%%%%%%%%%%%%%%%%%%%%%%%%%%%%%%%%%%%%%%%%%%%%%%%%%%%%
\title{Can quantum mechanics breed negative masses?}
%%%%%%%%%%%%%%%%%%%%%%%%%%%%%%%%%%%%%%%%%%%%%%%%%%%%%%%%%%%%%%

%%%%%%%%%%%%%%%%%%%%%%%%%%%%%%%%%%%%%%%%%%%%%%%%%%%%%%%%%%%%%%
\author{Bruno Arderucio Costa}
\email{Present address: Instituto de Ciencias Nucleares, Universidad Nacional Aut\'onoma de M\'exico, Mexico City, CDMX, 04510, Mexico. bruno.arderucio@correo.nucleares.unam.mx}
\affiliation{Instituto de F\'isica Te\'orica, Universidade Estadual Paulista, Rua Dr. Bento Teobaldo Ferraz, 271, 01140-070, S\~ao Paulo, Brazil}
%%%%%%%%%%%%%%%%%%%%%%%%%%%%%%%%%%%%%%%%%%%%%%%%%%%%%%%%%%%%%%

%%%%%%%%%%%%%%%%%%%%%%%%%%%%%%%%%%%%%%%%%%%%%%%%%%%%%%%%%%%%%%
\author{George E. A. Matsas}
\email{george.matsas@unesp.br}
\affiliation{Instituto de F\'isica Te\'orica, Universidade Estadual Paulista, Rua Dr. Bento Teobaldo Ferraz, 271, 01140-070, S\~ao Paulo, Brazil}
%%%%%%%%%%%%%%%%%%%%%%%%%%%%%%%%%%%%%%%%%%%%%%%%%%%%%%%%%%%%%%

%%%%%%%%%%%%%%%%%%%%%%%%%%%%%%%%%%%%%%%%%%%%%%%%%%%%%%%%%%%%%%
\begin{abstract}
    The Casimir effect realizes the existence of static negative energy densities in quantum field theory. We establish physically reasonable conditions for the non-negativity of the total mass of a Casimir apparatus held in equilibrium in the Minkowski background, irrespective of any condensed matter consideration. Specifically, the dynamical equilibrium requires the presence of additional matter to hold the system apart. As long as this extra matter satisfies the dominant energy condition, the mass of the combined system is positive. Thus, the very same reason why energy cannot travel backwards in time could be the underlying mechanism behind the positivity of the mass. We discuss the takeaways from the Casimir setting to more general circumstances.
\end{abstract}
%%%%%%%%%%%%%%%%%%%%%%%%%%%%%%%%%%%%%%%%%%%%%%%%%%%%%%%%%%%%%%

\maketitle

%%%%%%%%%%%%%%%%%%%%%%%%%%%%%%%%%%%%%%%%%%%%%%%%%%%%%%%%%%%%%%
\section{Introduction}
\label{Introduction}
%%%%%%%%%%%%%%%%%%%%%%%%%%%%%%%%%%%%%%%%%%%%%%%%%%%%%%%%%%%%%%

The general question of whether or not negative masses are allowed in general relativity has a long history. In 1957, Bondi~\cite{Bondi57} constructed a solution of Einstein's equations containing self-accelerating pairs of opposite-sign masses. A pictorial representation of his idea in the Newtonian regime is a pair of pointlike particles initially at rest with masses $m_+>0$ and $m_-<0$. The force on each particle points away from their partners (because $m_+m_-<0$) but the acceleration of $m_+$ is parallel to the force, while the acceleration of $m_-$ is antiparallel to it. The result is $m_-$ chasing after $m_+$, both accelerated.

As unacceptable as this may be, it is not so straightforward why some attractive fields could not generate enough negative potential energy to outweigh the positive mass of the bodily constituents. This conundrum is partially resolved by the positive-energy theorems~\cite{Schoen79, Schoen81, Witten81}, which rule out negative-mass systems as long as the initial conditions satisfy the dominant energy condition~(DEC) everywhere. However, since the DEC is not generally verified in the quantum realm, it is legitimate to wonder whether one can exploit quantum phenomena to allow for negative masses.

A simple example of a DEC-violating quantum system is the vacuum state between a pair of parallel Casimir plates. At first sight, it seems that if they were sufficiently light and close to each other, the overall mass could become negative.

It must be mentioned, however, that from an experimental perspective, actual endeavors to weigh the vacuum energy of a Casimir system would not reach anywhere close to this regime (see Ref.~\cite{Archimedes14} and references therein). The mass of a pair of plates set $1~\textup{~\AA}$ apart made of graphene with a surface density of $\sigma \approx 8 \times 10^{-7}~{\rm kg/m^2}$ exceeds the absolute value of the Casimir energy by almost a billion times. A more comprehensive analysis along these lines can be found in Ref.~\cite{Cerdonio15a}, where a microscopic model for the conducting plates is discussed. Albeit interesting, the reasoning above depends on (i)~condensed matter considerations, (ii)~the distance the plates are set apart, and also (iii)~the value of the product $c\hbar$. Were this product large enough, the Casimir energy could challenge the positivity of the overall system mass, depending on how this would impact the graphene's properties. In this paper, we wonder whether there exists a more fundamental reason why quantum mechanics could not contest the positivity of the mass. As we shall see, such reason may be the very same motive why energy cannot travel backward in time (codified in the dominant energy condition imposed on the classical struts).

From a conceptual perspective, Helfer concluded in Minkowski spacetime that to check for local DEC violations in a causal diamond between the plates, one should rely on clocks, whose own masses end up compensating for the negative vacuum energy density and ultimately leading to the overall compliance with the DEC~\cite{Helfer98}. Although interesting, the insertion of clocks perverts the original system. To avoid that, we shall directly probe the mass of the original system using gravity. This was also the route followed by Bekenstein~\cite{Bekenstein13}, who noted the plain but relevant fact that the Casimir effect would not exist in the absence of an auxiliary system in addition to the quantum field. In particular, (i)~plates made of conducting material are necessary to confine the Casimir electromagnetic vacuum and, most importantly (but not emphasized by Bekenstein), (ii)~struts are crucial to keeping the system in equilibrium. 

Bekenstein established a sufficient (albeit not necessary) requirement on the auxiliary system to guarantee the non-negativity of the overall mass of the whole structure. His analysis contemplates a wide class of fields, including the electromagnetic one. For that, he hypothesized that whichever additional matter composes the auxiliary system obeys the unorthodox subdominant trace energy condition (STEC):
\begin{equation}
\left|h_{ab}T_\text{AS}^{ab}\right|<T_\text{AS}^{ab}V_a V_b
    \label{stec}
\end{equation}
for any future-directed timelike vector field $V^a$, where $T_\text{AS}^{ab}$ is the energy-momentum tensor of the auxiliary system, $V^a V_a=-1$, and $h_{ab}\equiv g_{ab}+V_aV_b$ is the induced metric on the hypersurface orthogonal to $V^a$.

\begin{figure}[b]
    \centering
    \includegraphics[width=.75\columnwidth]{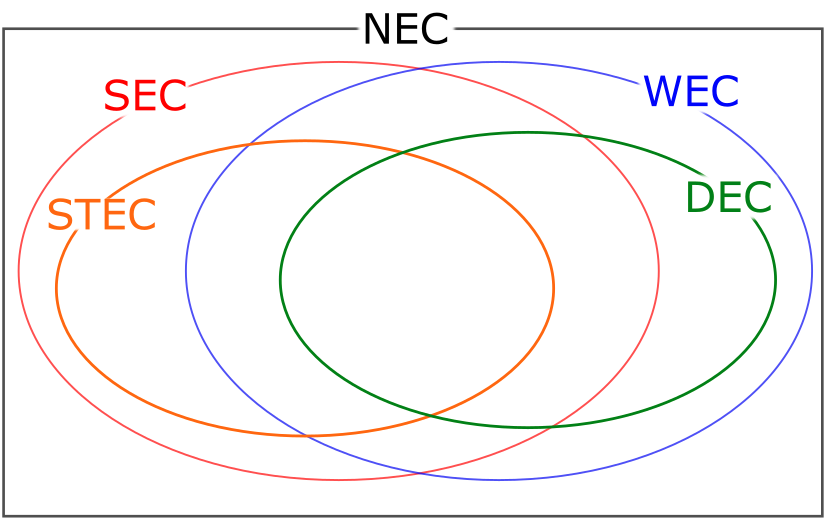}
    \caption{Venn diagram showing the implication relations between the various energy conditions. $A \subset B$ indicates $A\implies B$. WEC and~NEC  denote the weak and null energy conditions, respectively, included here for completeness. A perfect fluid with positive energy density, $\rho>0$, and pressure~$P$ in the interval $-\rho\leq P\leq-\rho/3$ lies within the DEC set but outside the SEC one.}
    \label{VennDiag}
\end{figure}

Bearing in mind that the positive-energy theorems often invoke the DEC and never the STEC, the need for the latter in Bekenstein's argument does not seem ``natural." In classical physics, the DEC, which demands that $-T^{ab} V_b$ is a future-directed nonspacelike vector for any future-directed timelike field $V^b$, guarantees that the energy propagation, associated with the energy-momentum tensor $T_{ab}$, is causal~\cite{HawkingEllis}.  In contrast, the physical interpretation of the~STEC is not as straightforward. Furthermore, it is not satisfied in standard models of dark energy, including the cosmological constant. For instance, a perfect fluid with positive energy density, $\rho>0$, in a four-dimensional spacetime with pressure $P$ in the interval $-\rho\leq P\leq-\rho/3$ satisfies the DEC but violates the strong energy condition~(SEC). One can see in the Venn diagram~\ref{VennDiag} that the~STEC is stronger than the SEC and, hence, by violating the latter, one is automatically violating the former: 
$$
{\rm STEC} \implies {\rm SEC},\quad
\neg{\rm SEC} \implies \neg{\rm STEC}.
$$ 
Here, we look for more natural physical constraints or processes that could emulate a ``cosmic-weight watcher'', which should forbid compact-core isolated systems in asymptotically flat spacetimes to possess negative masses.

The paper is organized as follows. In Sec.~\ref{Mass} we review some known facts about the Komar mass, which, nevertheless, are relevant to the present paper. In Sec.~\ref{The basic example} we analyze the parallel-plate Casimir system in the vacuum and verify that it suffices that the classical matter needed to sustain the system obeys the DEC for the positivity of the total mass. In Sec.~\ref{More general cases} we show that our conclusions continue to hold for other static states. Finally, in Sec.~\ref{Discussion} we present our final remarks. We hereafter adopt natural units $G=c=\hbar=1$ and metric signature $(-,+,+,+)$, unless stated otherwise.

%%%%%%%%%%%%%%%%%%%%%%%%%%%%%%%%%%%%%%%%%%%%%%%%%%%%%%%%%%%%%%
\section{Mass}
\label{Mass}
%%%%%%%%%%%%%%%%%%%%%%%%%%%%%%%%%%%%%%%%%%%%%%%%%%%%%%%%%%%%%%

The absolute value of the mass of an isolated system in an asymptotically flat stationary spacetime is ultimately fixed by the gravitational field and can be assessed by distant test particles. We assume that they react to the spacetime as ruled by the semiclassical Einstein equations
\begin{equation}
G_{ab} 
= 
{8 \pi} 
\left \langle T_{ab} \right \rangle,
\label{SEE}
\end{equation}
where 
\begin{equation}
\left \langle T^{ab} \right \rangle 
\equiv 
\left\langle \hat{T}_\text{EM}^{ab}\right\rangle
+
T_\text{AS}^{ab}
\label{totalT}
\end{equation}
encompasses the regularized vacuum expectation value of the energy-momentum tensor of the quantum fields, $\left\langle\hat{T}_\text{EM}^{ab}\right\rangle$, 
{\em e.g.}, the electromagnetic one, and of the auxiliary system, 
$T_\text{AS}^{ab}$, 
{\em e.g.}, plates and struts. It is only the combination of both  $\left\langle\hat{T}_\text{EM}^{ab}\right\rangle$ and $T_\text{AS}^{ab}$ that is expected to be conserved:
\begin{equation}
\nabla_a \left\langle T^{ab} \right\rangle =0.
\label{conservation}
\end{equation}

Although Eq.~\eqref{SEE} demands, in general, that the stress-energy tensor has small dispersion~\cite{Kuo93}, this equation should still provide a good asymptotic description of the spacetime for our stationary compact-core systems, however complex they might be, because they all look pointlike when probed from sufficiently large distances. We will discuss this in greater detail later in the paper.

In asymptotically flat stationary spacetimes with a timelike Killing field $\xi^a$, this notion is captured in Komar's formula, which can be written as an integral over a topological 2-sphere $S$ at infinity:
\begin{equation}
M=-\frac{1}{8\pi}\oint_{S} \ud S_{[ab]}\nabla^a\xi^b.
    \label{KomarDef}
\end{equation}
Grounded on observations, it is reasonable to expect that all isolated compact-core systems, for which Eq.~\eqref{KomarDef} is well defined, possess non-negative masses irrespective of whether their composition is dominated by quantum or classical matter. 

In nearly flat backgrounds, Eq.~\eqref{KomarDef} can be cast in a simpler form. For that, let us express it in terms of an integral over a spacelike hypersurface~$\Sigma$ bounded by~$S$, as usual~\cite{Wald84}. Assuming the semiclassical Einstein equations~\eqref{SEE} the result is
\begin{equation}
    M=2\int_\Sigma\left(\left\langle T_{ab} \right\rangle -\frac{1}{2}g_{ab} \left\langle T_c^{\; c} \right\rangle \right)\xi^an^b\ud\Sigma,
    \label{KomarBulk}
\end{equation}
where  $n^a$ is a future-directed timelike unit vector orthogonal to $\Sigma$. 

Next, let us assume that the system is not only stationary but also static and the spacetime is nearly flat. Then, we can replace $g_{ab}$ in Eq.~\eqref{KomarBulk} by the Minkowski metric $\eta_{ab}$ to obtain $M$ in the leading order. In Cartesian coordinates, $x^\mu=(t,x,y,z)$, according to which the Minkowski line element is 
$$
\ud s^2=-\ud t^2+\ud x^2+\ud y^2+\ud z^2,
$$
our static system satisfies 
$$ 
\partial_0 \langle T^{\mu \nu} \rangle=0,  \quad \langle T^{0i}\rangle =0,\;\;  i=1,2,3,
$$
and Eq.~\eqref{conservation} reduces to
\begin{equation}
\partial_i \langle T^{i j} \rangle =0.
\label{aux2}
\end{equation}
In these coordinates, $\xi^\mu=n^\mu=(1,0,0,0)$, and Eq.~\eqref{KomarBulk} simplifies to
\begin{eqnarray}
    M 
    &=& - \int \ud^3 x 
    \left( 
          \langle T_{0}^{\; 0} \rangle  
         + \langle T_i^{\; i} \rangle 
    \right) \nonumber \\
    &=&
    - \int \ud^3 x 
    \left( 
           \langle T_{0}^{\; 0} \rangle  
          + \langle T_i^{\; i} \rangle 
    \right) 
    +\oint_S \ud S_j\; x^i \langle T_{i}^{\; j}\rangle
    \nonumber \\
     &=&
    - \int \ud^3 x \; 
          \langle T_{0}^{\; 0} \rangle  
    +\int \ud^3 x\; x^i \; \partial_j  \langle T_{ i}^{\;j}  \rangle
    \nonumber \\
    &=&
    \int \ud^3 x \; 
          \langle T_{00} \rangle  ,
    \label{KomarMinkowski}
\end{eqnarray}
wherein the second line we have added an integral over an arbitrarily large 2-sphere (which identically vanishes provided that $\langle T_{ i}^{\;j} \rangle$ decays faster than $r^{-3}$ at infinity, as verified in most cases of interest) and the last two lines, we have used the Stokes's theorem and Eq.~\eqref{aux2}, respectively. Equation~\eqref{KomarMinkowski} conveys a well-known result, namely, that in flat spacetime the mass~$M$ as probed by asymptotic test masses, corresponds to the integral of the energy density, which, in our case, encompasses both classical matter and quantum fields. Furthermore, Eq.~\eqref{KomarMinkowski} is in agreement with Ref.~\cite{Archimedes14}, where the authors conclude that the vacuum-buoyancy force on a Casimir cavity in a weak gravitational field is $F= E_{\rm cas} g$ with $E_{\rm cas}$ being the Casimir energy and $g$ the gravitational acceleration.

In the next section, we analyze the paradigmatic parallel-plate Casimir system and identify the physical ingredient that prevents the total mass from becoming negative. In summary, we establish that a sufficient condition for the positivity of the mass of the system comprising the Casimir field, parallel plates, and struts is that the latter, which sustain the system in equilibrium, obey the DEC.

We also remark that despite the local fluctuations in the stress-energy tensor in the Casimir vacuum being of the same order of magnitude as 
$\left \langle \hat{T}_{00}^{\rm EM } \right \rangle$~\cite{Kuo93}, 
the vacuum state is, by definition, an eigenstate of the Hamiltonian operator:
$$
\hat{H} = \int \hat{T}_{00}^{\rm EM}\  \ud^3x
$$ 
and, thus,  the total energy of the Casimir vacuum does not fluctuate at all. This reinforces our reasoning in favor of Eq.~\eqref{KomarMinkowski}, derived from Eq.~(\ref{SEE}), as a reliable semiclassical approximation for how asymptotic test particles feel the presence of the total Casimir apparatus.

%%%%%%%%%%%%%%%%%%%%%%%%%%%%%%%%%%%%%%%%%%%%%%%%%%%%%%%%%%%%%%
\section{Casimir Vacuum}
\label{The basic example}
%%%%%%%%%%%%%%%%%%%%%%%%%%%%%%%%%%%%%%%%%%%%%%%%%%%%%%%%%%%%%%

Not all Casimir geometries threaten the general expectation of non-negative total masses. The Casimir effect yields a positive energy
$$
\int_{\mathbb R^3}\ud^3x\ \left\langle \hat{T}^{\rm EM}_{00}\right\rangle >0
$$
when the boundary conditions take place on the surface of a sphere~\cite{Mostepanenko97, Milton04}. Therefore, we dwell on the original, parallel-plate geometry, the quintessence of the configurations that require scrutiny.

Let us start by considering the idealized case of a pair of infinitely thin conducting plates at rest located at $z=0$ and $z=a$. The proper boundary conditions on the electromagnetic field $F_{ab}$ can be cast as
\begin{eqnarray}
    F^{0x}(t,x,y,0)=F^{0x}(t,x,y,a) &=& 0, \label{bc1}\\
    F^{0y}(t,x,y,0)=F^{0y}(t,x,y,a) &=& 0, \label{bc2}\\
    F^{xy}(t,x,y,0)=F^{xy}(t,x,y,a) &=& 0, \label{bc3}
\end{eqnarray}
where the first two lines impose that the tangential components of the electric field vanish on the plates and the third line enforces a vanishing orthogonal component of the magnetic field on the same surfaces.

The (regularized) vacuum expectation value of the stress-energy tensor can be found using standard techniques~\cite{Mostepanenko97, Birrell84, Milton08}:
\begin{equation}
    \left\langle \hat{T}^{\rm EM}_{\mu\nu}\right\rangle=\frac{\pi^2}{720a^4}\ \mathrm{diag}(-1,1,1,-3) [\theta(z)-\theta(z-a)].
    \label{confscalar}
\end{equation}
However, as stated above, the electromagnetic field cannot exist in this state in isolation because  
$$
\nabla_a\left\langle \hat{T}_{\rm EM}^{ab}\right\rangle\neq 0
$$
at $z=0$ and $z=a$. The conservation relation~\eqref{conservation} demands the presence of a certain additional system, whose energy-momentum tensor $T^{ab}_\text{AS}$ must satisfy
\begin{eqnarray}
    \partial_\mu T^{\mu \nu}_\text{AS}
    &=&-\partial_\mu\left\langle \hat{T}^{\mu \nu}_{\rm EM}\right\rangle
    \nonumber \\
    &=&
    \frac{\pi^2 }{240 a^4}\eta^{z \nu}[\delta(z)-\delta(z-a)].    
    \label{divstruts}
\end{eqnarray}
%Any system consisting of the electromagnetic field and an additional energy-momentum tensor governed by Eq.~\eqref{divstruts} is automatically in dynamical equilibrium.

We seek regular, bounded solutions to Eq.~\eqref{divstruts} that vanish at $z=\pm\infty$. A homogeneous solution on the $xy$ plane for $T_\text{AS}^{zz}$ is
\begin{equation}
T^{zz}_\text{AS}=\frac{\pi^2}{240 a^4}[\theta(z)-\theta(z-a)].
    \label{struts}
\end{equation}
Equation~\eqref{struts} must encompass both the conducting plates at $z=0$, $z=a$, and some material layer functioning as struts, which must be there to balance out the attractive force between the plates. Here, we have assumed that the electromagnetic field does not interact significantly with the layer that fills the region between the plates. This assumption is naturally vindicated in physical situations where finite plates are held apart by thin struts. Nevertheless, not even infinite struts as above cause concern. For isotropic nondispersive struts with permittivity $\epsilon$ and permeability $\mu$, e.g., the corresponding vacuum expectation value of the energy-momentum tensor, can be shown to differ from Eq.~\eqref{confscalar} by the constant factor $1/\sqrt{\epsilon \mu}$~\cite{Brevik01}. Since our arguments rely only on the ratio $\left\langle\hat T^{EM}_{00}\right\rangle/\left\langle\hat T^{EM}_{zz}\right\rangle$, struts composed of such materials do not jeopardize our conclusions.
%This corresponds to replacing $c=1$ in Eq.~\eqref{confscalar} with the speed of light in the medium. Since none of our conclusions depend on the magnitude of the speed of light, our conclusions are not jeopardized by this assumption.}

If the struts are made of classical matter, it is very reasonable to impose the DEC on its energy-momentum tensor. Equation~\eqref{struts} then entails 
\begin{equation}
T^{00}_\text{AS}\geq\left|T^{zz}_\text{AS}\right|=\frac{\pi^2}{240 a^4}[\theta(z)-\theta(z-a)].
\label{DECcon}
\end{equation}
Now, using Eqs.~\eqref{confscalar} and~\eqref{DECcon}, we see that the DEC is sufficient to guarantee that the mass of the system~\eqref{KomarMinkowski} is positive:
\begin{equation}
    M= 
    \int\ud^3x 
    \left( \left\langle \hat{T}_{00}^{\rm EM}\right\rangle + T_{00}^\text{AS} \right) \geq \frac{\pi^2 A}{360 a^3},
    \label{positivemass}
\end{equation}
where $A$ is the surface area of the plates. We can already see that the struts are the central elements responsible for the positivity of the mass.     

The result obtained thus far assumed arbitrarily thin and large conducting plates held in place by classical matter occupying the whole space between the plates, as per Eq.~\eqref{struts}. Nevertheless, this is no more than an idealization for the actual physical situation, in which the plates are much larger than the separation between them. Indeed, material plates with constant surface mass density~$\sigma$ and area
$$
A \gtrsim c^4/(G^2 \sigma^2)
$$ 
are expected to collapse under their weight according to the hoop conjecture~\cite{MisnerThorneWheeler, Flanagan91}. Thus, we must move on and consider finite plates. (We refer the reader to Refs.~\cite{Schwinger78--Klimchitskaya09} and references therein for discussions on the Casimir effect under more realistic conditions.)

Instead of solving the whole problem for finite plates with edge effects, which would be a superlative undertaking, we exploit that the system is in equilibrium and satisfies Eq.~\eqref{conservation}, and the struts must compensate the attractive force between the plates, as in the infinite-plate case.
Finite parallel plates can be held apart by thin struts, covering a transverse area $\delta A<A$. %Finite parallel plates can be held apart by thin struts, covering a transverse area $\delta A<A$. Instead of solving the whole problem with edge effects, which would be a superlative undertaking, we can still use the infinite plate as a model and refrain from summoning that $\nabla_b\left(\langle T^{ab}\rangle+T^{ab}_\text{struts}\right)=0$ pointwise, but merely that the struts compensate the attractive force between the plates. 
Per unit area, this force is 
$\left\langle \hat{T}_{zz}^{\rm EM}\right\rangle$~\cite{footnote}. Hence, the struts must provide a pressure
\begin{eqnarray}
T_{zz}^{\rm struts} 
&=& -\left\langle \hat{T}_{zz}^{\rm EM} \right\rangle \frac{A}{\delta A}
\label{pressurestrutspressurefield}\\
&=& \frac{\pi^2}{240 \, a^4} \frac{A}{\delta A},
\label{pressurestrutspressurefield2}
\end{eqnarray}
where we have used Eq.~\eqref{confscalar} as a good estimation of $\left\langle \hat{T}_{zz}^{\rm EM} \right\rangle$ for plates close enough to each other in comparison to their size.

As in the infinite-plate case, if the struts comply with the DEC, they must possess an energy density of at least the same magnitude, {\em i.e.},
\begin{equation}
T_{00}^{\rm struts} \geq |T_{zz}^{\rm struts}|.
\label{DECstruts}
\end{equation}

The spatial integral of the energy density~\eqref{DECstruts} is independent of $\delta A$, meaning that no matter how thin the struts supporting the plates are, {\em the mass of the system consisting of the field and the struts alone is positive}:
\begin{eqnarray}
M &=& 
    \int\ud^3x 
    \left( \left\langle \hat{T}_{00}^{\rm EM}\right\rangle + T_{00}^\text{struts} \right) 
\label{positivemass2c}\\
&\geq&
\int\ud^3x 
    \left( 
      \left\langle \hat{T}_{00}^{\rm EM}\right\rangle 
    + \left| \left\langle 
    \hat{T}_{zz}^{\rm EM} 
    \right\rangle \right| 
    \right)
    \label{positivemass2b}\\
&\geq&
\frac{\pi^2 A}{360 a^3},
    \label{positivemass2}
\end{eqnarray}
where in the second and third lines we  have used Eqs.~\eqref{pressurestrutspressurefield},~\eqref{DECstruts} and~Eqs.~\eqref{confscalar},~\eqref{pressurestrutspressurefield2}, respectively.  Thus, although for practical reasons the vacuum Archimedes effect cannot be  measured in the laboratory by comparing the weight of the assembled cavity with the sum of the weights of its parts~\cite{Archimedes14}, the mass of the struts alone should weigh more than the absolute value of the Casimir energy. 

Numerical analyses provide an estimate for the systematic errors when approximating the force between a pair of finite plates using the expression for the infinite planes. The results of Ref.~\cite{Gies06} suggest that the energy densities do not change much as a result of the edge effects, but they spread out beyond the limits of the plates. Consequently, the correction could be incorporated as an effective increase in the area $A$~\cite{Gies06} in Eq.~\eqref{pressurestrutspressurefield} and in the integration region of Eq.~\eqref{positivemass2c}, so that the result Eq.~\eqref{positivemass2} would remain positive.

The analysis above is unaltered if we replace the electromagnetic field with any other conformal field, since its stress-energy tensor only differs from Eq.~\eqref{confscalar} by a constant factor~\cite{Brown69, deWitt75}.

Finally, we emphasize that we have made no hypotheses about the physics of the materials that constitute the plates. They can be made arbitrarily light, for DEC-abiding struts alone guarantee the result~\eqref{positivemass2}.

To help conceive the effect of a more realistic model for the conductivity of the plates that is not captured in the boundary conditions~\eqref{bc1}--\eqref{bc3}, we can study the dependence of the electromagnetic stress-energy tensor with an ultraviolet cutoff. This has been worked out in Ref.~\cite{Hagen01}. One can see from their Eqs.~(9) and (8) that the introduction of finite cutoffs is to increase the electromagnetic energy density and the pressure, respectively. Additionally, the effect on the energy density is bigger than the effect on the pressure. Hence, a repetition of our argument reveals that the presence of the cutoffs makes the configuration not as arduous to the cosmic-weight watcher.

At this point, we conjecture that all that the cosmic-weight watcher must impose on the classical matter to safeguard the positivity of the overall mass of any Casimir-like configuration is the DEC. In the next section, we shall see that this conjecture survives when the electromagnetic vacuum is replaced with an excited state.

%%%%%%%%%%%%%%%%%%%%%%%%%%%%%%%%%%%%%%%%%%%%%%%%%%%%%%%%%%%%%%
\section{Excited States}
\label{More general cases}
%%%%%%%%%%%%%%%%%%%%%%%%%%%%%%%%%%%%%%%%%%%%%%%%%%%%%%%%%%%%%%

Hitherto, we only considered one state of the field, the Casimir vacuum. If the field is no longer prepared in this state, it is necessary to specify in which state the expectation values are computed. For this reason, we henceforth denote the Casimir vacuum by $\mathbf C$ and a generic stationary Hadamard state by $\Psi$. Our aim, now, is to set a bound on 
\begin{equation}
    M= 
    \int\ud^3x 
    \left( \left\langle \hat{T}_{00}^{\rm EM}\right\rangle_\Psi + T_{00}^\text{struts} \right)
    \label{M_Psi}
\end{equation}
analogous to Eqs.~\eqref{positivemass2c}--\eqref{positivemass2}. Imposing the DEC, as before,  on the struts and recalling that the pressure on them is just the necessary one to keep the system in equilibrium, we write
\begin{equation}
    M\geq \int\ud^3x\ \left(\left\langle\hat T_{00}^\text{EM}\right\rangle_\Psi+\left|\left\langle\hat T_{zz}^\text{EM}\right\rangle_\Psi\right|\right).
    \label{totmassmod}
\end{equation}
Clearly, $M>0$ if  
$$
\left\langle\hat T_{00}^\text{EM}\right\rangle_\Psi \geq 0,
$$ 
where we assume that  $\left|\left\langle\hat T_{zz}^\text{EM}\right\rangle_\Psi\right|$ 
is strictly positive. Hence, the only case which requires a closer examination is the one for which, as for the Casimir vacuum, 
\begin{equation}
\left\langle\hat T_{00}^\text{EM}\right\rangle_\Psi<0.
\label{col0}
\end{equation}

In the Appendix, we derive the following inequality valid for any Hadamard state $\Psi$ for which the expectation value of the stress-energy tensor does not depend on $t$:
\begin{equation}
\left\langle\hat T_{00}^\text{EM}\right\rangle_\Psi-\left\langle\hat T_{zz}^\text{EM}\right\rangle_\Psi-\left(\left\langle\hat T_{00}^\text{EM}\right\rangle_\mathbf{C}-\left\langle\hat T_{zz}^\text{EM}\right\rangle_\mathbf{C}\right)\geq 0.
\label{col1}
\end{equation}
Now, from Eq.~\eqref{confscalar} we know $\left\langle\hat T_{zz}^\text{EM}\right\rangle_\mathbf{C}=3\left\langle\hat T_{00}^\text{EM}\right\rangle_\mathbf{C}$. Substituting into Eq.~\eqref{col1}, we have
\begin{equation}
    \left\langle\hat T_{zz}^\text{EM}\right\rangle_\Psi
    \leq  
    2\left\langle\hat T_{00}^\text{EM}\right\rangle_\mathbf{C}+\left\langle\hat T_{00}^\text{EM}\right\rangle_\Psi
    < 0,
    \label{ppsi}
\end{equation}
where we have used Eqs.~\eqref{confscalar} and~\eqref{col0}. Next, using Eq.~\eqref{ppsi}, we rewrite Eq.~\eqref{totmassmod} as
\[M \geq \int\ud^3x\ \left(\left\langle\hat T_{00}^\text{EM}\right\rangle_\Psi-\left\langle\hat T_{zz}^\text{EM}\right\rangle_\Psi\right),\]
or, after applying Eq.~\eqref{col1} once again,
\begin{equation}
M
\geq 
\frac{\pi^2 A}{360 a^3},
\label{Mpsi}
\end{equation}
as we intended to show.

%%%%%%%%%%%%%%%%%%%%%%%%%%%%%%%%%%%%%%%%%%%%%%%%%%%%%%%%%%%%%%
\section{Conclusions}
\label{Discussion}

The existence of compact-core isolated systems in asymptotically flat spacetimes with negative mass would lead to runaway solutions. Naked singularities with angular momentum~$J$ and mass $M<0$ as described by the line element 
\begin{eqnarray}
\ud s^2
&=&
-\left(1- \frac{2M}{r} \right)\ud t^2
+\left(1+ \frac{2M}{r} \right) \left( \ud r^2+ r^2 \ud\Omega^2 \right) 
\nonumber \\
&-& 4 \frac{J}{r} (\sin \theta)^2 \ud t \ud\varphi
\label{poisson}
\end{eqnarray}
could be ruled out by evoking the cosmic censorship conjecture but it is not obvious how negative-mass regular systems with asymptotic line element as given by Eq.~\eqref{poisson} would be ruled out from nature in practice. This dilemma is alleviated  by the positive-energy theorems~\cite{Schoen79--Witten81},  which rule out negative-mass systems as long as the initial conditions satisfy the DEC everywhere.  However, they are silent when the DEC is violated as it is common in the quantum realm. This observation motivated our inquiry into what physical conditions a cosmic-weight watcher is likely to resort to when forbidding the existence of negative-mass systems.

We have analyzed this problem in the Casimir system of idealized parallel plates, a paradigmatic instance where the expectation value of the energy-momentum tensor violates the DEC.  We have shown that the very imposition of the DEC on the classical struts, which make sure they comply with causality, is enough to guarantee the positivity of the overall mass. The conclusion above persists when the electromagnetic field is replaced by other conformal fields, and the Casimir vacuum $\mathbf{C}$ by more general Hadamard states~$\Psi$. That appeases concerns about self-accelerating pairs of Casimir systems and hints at why negative masses have never been observed for small systems dictated by quantum mechanics. 

The idealized configuration described in the paper was chosen for being, seemingly, the most challenging limiting case. For instance, thick plates would add positive energy to the system, and finite conductivity should make the plates more transparent to high-frequency vacuum modes mitigating quantum effects. Also, because the Casimir energy between the plates goes as $-1/a^3$, the situation that challenges the mass positivity the most is when the plates are close to each other, where edge effects matter the least. Nevertheless, it is not as obvious whether an ``almost-plane'' geometry could be more challenging than the usual ``perfectly plane'' configuration. Although we do not have a clear-cut answer to this, we recall that the DEC on the struts is sufficient, not necessary, to ensure the non-negativity of the mass of the system (the lower bounds~(\ref{positivemass}) and (\ref{positivemass2c}) are strictly positive). Thus, the DEC leaves some room to guarantee positivity of the mass in situations somewhat more challenging than that one posed by the ideal metallic plates. We hope this is enough to cover any more challenging configuration (if any).  

Although our analyses were restricted to a flat spacetime, we expect their upshot to remain valid for general well-behaved globally hyperbolic asymptotically flat spacetimes. The cosmic-weight watcher must rule out from nature regular asymptotically flat stationary solutions of Einstein's equations with $M<0$. For example, a Morris-Thorne wormhole, described by the metric 
$$\ud s^2=
-\ue^{2\Phi(r)}\ud t^2+\frac{\ud r^2}{1-b(r)/r}+r^2\ud\Omega^2,
$$ 
where $b(r)$ is the shape function, violates the DEC~\cite{Morris88, HawkingEllis}, thereby escaping the positive-energy theorems. Nevertheless, if the wormhole is surrounded by vacuum, its Komar mass is evaluated from Eq.~\eqref{KomarDef} and reads $$M=\lim_{r\to\infty}\frac{1}{2}\ \frac{b(r)}{1-b(r)/r}.$$  Hence, the mass is manifestly non-negative when the solution is regular, i.e., when $0\leq b(r)< r$ everywhere. Even though we are not claiming that wormhole solutions are physical if they pass the cosmic-weight watcher benchmark, this example illustrates why we must insist on our assumptions of regularity and ``core compactness." The latter, in this particular case, is implemented by requiring the matter distribution to vanish ($T_{ab}=0$) outside of a bounded set.

%When gravity is negligible, all forces on a finite system must eventually balance out. This fact, manifest in Eq.~\eqref{Bruno}, entails that the Casimir plates and field must coexist with struts. However, gravity could, in principle, modify this conclusion. In the original system without the struts, gravity acts repulsively (as can be seen from Raychaudhuri's equation). Could it equilibrate the plates for free, jeopardizing our conclusions?

%Our perturbative approach suggests that this is not so. The full Komar formula~\eqref{KomarBulk} includes contributions from the positive net pressure necessary to prevent a generic body from collapsing under its own weight. In the first order, the deviations from the Minkowski background produce Eq.~\eqref{usual}, and struts must be added. In the next leading order, the effects of the struts add even more positively on the integrand of Eq.~\eqref{KomarBulk}.

%Yet, to avoid known counterexamples of geometries with negative masses, further assumptions must be made. The spacetime must be free of naked singularities, and, as we shall argue in a particular case, it must be a solution to Einstein's equations

\acknowledgements
The authors are thankful to Steve Fulling and Kim Milton for reading the paper and correcting a statement concerning Eq.~\eqref{struts}. B.A.C. and G.E.A.M. were fully and partially funded by Conselho Nacional de Desenvolvimento Cient\'ifico e Tecnol\'ogico (CNPq) under Grants No. 162288/2020-4 and No. 301544/2018-2, respectively.

\appendix
\section{PROOF OF EQ.~\eqref{col1}}
To prove Eq.~\eqref{col1} of the main text, we need to introduce some notation. We quantize the electromagnetic field $F_{\mu\nu}=\partial_\mu A_\nu-\partial_\nu A_\mu$ in the Coulomb gauge, $A^0=0$ and $\partial_i A^i=0$. The quantized potentials are expanded in terms of a set of positive-norm solutions $\mathcal A^\mu_{\lambda\vec k}(x)$ of Maxwell's equations with the boundary conditions~(\ref{bc1})--(\ref{bc3}):
\begin{equation}
\hat A^\mu
=
\sum_{\lambda=1}^2\int\ud^3\vec k\ 
\left[
\mathcal A^\mu_{\lambda\vec k}(x)\hat a_{\lambda\vec k}
+
\overline{\mathcal A^\mu_{\lambda\vec k}}(x)\hat a^\dagger_{\lambda\vec k}
\right],
    \label{modeexp}
\end{equation}
where the operators $\hat a_{\lambda\vec k}$ annihilate $\mathbf C$ for every $\vec k$ and $\lambda$. Despite knowing that $k_z$ is only allowed to assume a discrete set of values, we write $\int\ud^3\vec k$ instead of $\iint\ud k_x\ud k_y\sum k_z$ to avoid overloading our notation. To obtain Eq.~\eqref{col1}, we first note that 
$$
\hat T_{00}^\text{EM}-\hat T_{zz}^\text{EM}=\hat E_z^2+\hat B_z^2,
$$ 
where, in the Coulomb gauge, 
\begin{equation*}
\hat E_i=-\partial_0 \hat A_i\quad 
{\rm and}
\quad \hat B_i=\varepsilon_i^{\ jk}\partial_j \hat A_k
\end{equation*}
are operators representing the electric and magnetic field vectors.

Using the mode expansion~\eqref{modeexp} and the canonical commutation relations 
$$
\left[\hat a_{\lambda\vec k},\hat a^\dagger_{\lambda^\prime\vec k^\prime}\right]=\delta_{\lambda\lambda^\prime}\delta^{(3)}(\vec k-\vec k^\prime),
$$ 
we obtain
\begin{widetext}
\begin{eqnarray}
    \left\langle\hat T_{00}^\text{EM}\right\rangle_\Psi
    -
    \left\langle \hat T_{zz}^\text{EM}\right\rangle_\Psi
    &=&
    \sum_\lambda\int\ud^3\vec k\ 
    \left[\overline{\mathcal E_z^{\lambda \vec k}} 
                    \mathcal E_z^{\lambda \vec k} 
               +\overline{\mathcal B_z^{\lambda \vec k}} 
                    \mathcal B_z^{\lambda \vec k}     
         \right]
+ 
    2\Re\sum_{\lambda\lambda^\prime}\int\ud^3\vec k\int\ud^3\vec k^\prime
    \left\{\left[
    \overline{\mathcal E_z^{\lambda^\prime \vec k^\prime}} 
              \mathcal E_z^{\lambda \vec k}
             +\overline{\mathcal B_z^{\lambda^\prime \vec k^\prime}} 
              \mathcal B_z^{\lambda \vec k}
           \right]
           \left\langle 
             \hat a^\dagger_{\lambda\vec k}
             \; \hat a_{\lambda' \vec k'}
           \right\rangle_\Psi
          \right.
\nonumber \\  
     &+& \left.
    \left[
    \mathcal E_z^{\lambda^\prime \vec k^\prime} 
    \mathcal E_z^{\lambda \vec k}
    +\mathcal B_z^{\lambda^\prime\vec k^\prime} 
     \mathcal B_z^{\lambda \vec k}  \right]
     \left\langle
     \hat a_{\lambda\vec k}\hat a_{\lambda^\prime\vec k^\prime}\right\rangle_\Psi\right\},
    \label{pres}
\end{eqnarray}
\end{widetext}
where 
\begin{equation}
\mathcal E_i^{\lambda \vec k} 
\equiv 
\partial_0\mathcal A_i^{\lambda \vec k} (x)\quad 
 {\rm and} \quad
\mathcal B_i^{\lambda \vec k} (x) 
\equiv 
\varepsilon_i^{\ jk}\partial_j \mathcal A_k^{\lambda \vec k} (x)
\label{ebcoulomb}
\end{equation}
are the electric and magnetic field vectors obtained from the potential $\mathcal A^\mu$. We now apply the following positivity result by Pfenning~\cite{Pfenning01} valid for any Hadarmard state $\Psi$ of the electromagnetic field, for any $n\times n$ real symmetric positive semidefinite matrix $\mathbf M$, and for any complex $n\times1$ matrix $\mathbf P_{\lambda \vec{k}}$ with Hermitian conjugate 
$\mathbf P^\dagger_{\lambda \vec{k}}$:
\begin{widetext}
\begin{eqnarray}
&\Re &
\sum_{\lambda\lambda^\prime}\int\ud^3\vec k\int\ud^3\vec{k^\prime}
\left\{\tilde f(\omega^\prime-\omega)
\left\langle 
\hat a_{\lambda\vec k}^\dagger 
\hat a_{\lambda^\prime\vec{k^\prime}}
\right\rangle_\Psi
\pm
\tilde f(\omega+\omega^\prime)\left\langle \hat a_{\lambda\vec k}\hat a_{\lambda^\prime\vec{k^\prime}}\right\rangle_\Psi
 \right\} \mathbf P^\dagger_{\lambda \vec k}
\mathbf M
\mathbf P_{\lambda^\prime \vec k^\prime}
\nonumber \\
&\geq&
- \frac{1}{2\pi}\int_0^\infty  \ud\alpha \sum_\lambda\int\ud^3\vec k\ \left|\widetilde{f^{\frac{1}{2}}}(\omega+\alpha)\right|^2 \mathbf P^\dagger_{\lambda \vec k} \mathbf M 
\mathbf P_{\lambda \vec k},
    \label{EMQEI}
\end{eqnarray}
\end{widetext}
where 
$\omega^2\equiv\vec k^2$, 
$\omega^\prime \equiv {\vec k}^{\prime 2}$, 
and 
$\tilde f\equiv\int_{-\infty}^\infty f(t)\ue^{-\ui\omega t}\ud t$ is the Fourier transform of any real, infinite-differentiable, square-integrable test function $f$. In general, $f$ serves as a ``sampling function'' but since we are interested in static observables in static spacetimes, we can take
\begin{equation}
    f(t)=\frac{t_0}{\pi(t_0^2+t^2)} %\qquad t_0\to\infty.
    \label{Lorentzian}
\end{equation}
and eventually take the optimal limit in which the sampling time $t_0$ spreads out indefinitely $t_0\to\infty$.

To set a bound on Eq.~\eqref{pres} using Eq.~\eqref{EMQEI}, we first subtract from Eq.~\eqref{pres} the analogous equation for the state $\mathbf C$ so that the first term drops. Now, $\mathcal{A}(\lambda,\vec k;x)$ has a time dependence of $\ue^{-\ui\omega t}$ and the vectors $\mathcal E^i$ and $\mathcal B^i$ inherit this dependence. Therefore, after smearing the resulting equation with the test function $f$ in Eq.~\eqref{Lorentzian}, the difference
\begin{widetext}
\begin{equation}
\mathcal D\equiv\int_{-\infty}^\infty\left[\left\langle\hat T_{00}^\text{EM}\right\rangle_\Psi-\left\langle\hat T_{zz}^\text{EM}\right\rangle_\Psi-\left(\left\langle\hat T_{00}^\text{EM}\right\rangle_\mathbf{C}-\left\langle\hat T_{zz}^\text{EM}\right\rangle_\mathbf{C}\right)\right]f(t)\ud t
\label{auxa1}
\end{equation}
has the same form as two copies of the left-hand side of Eq.~\eqref{EMQEI} with $M_{ij}=\delta_{iz}\delta_{jz}$, one for $P^i=\mathcal E^i$ and another for $P^i=\mathcal B^i$. Thus, we conclude that, for any static state $\Psi$,

%Owing to its technical simplicity, we begin with the minimally coupled scalar field. We use the quantum energy inequalities (reviewed, e.g., in ref.~\cite{Fewster2017}) for the Casimir setting in Minkowski spacetime and the vacuum $\mathbf C$ as a reference state.

%With this choice, applying Eq.~(\ref{QEI}) consecutively for $Q=\frac{\partial}{\partial x}$, $Q=\frac{\partial}{\partial y}$ and $Q=m$ and summing we obtain

%Subtracting from Eq.~\eqref{pres} the same equation for the state $\mathbf C$, the first term drops. Now $\mathcal A(\lambda,\vec k;x)$ has time dependence of $\ue^{-\ui\omega t}$, and the vectors $\mathcal E^i$ and $\mathcal B^i$ inherit this dependence. Therefore, after smearing the resulting equation with the test function~\eqref{Lorentzian} and integrating, the difference $$\int\left[\langle T_{00}\rangle_\Psi-\langle T_{zz}\rangle_\Psi-\left(\langle T_{00}\rangle_\mathbf{C}-\langle T_{zz}\rangle_\mathbf{C}\right)\right]|f(t)|^2\ud t$$ has the same form as two copies of the left-hand side of Eq.~\eqref{EMQEI} with $M^{ij}=\delta^{iz}$, one for $P^i=\mathcal E^i$ and another for $P^i=\mathcal B^i$. Thus, we conclude that, for any static state $\Psi$,

\begin{equation}
%\int\left[\left\langle\hat T_{00}^{EM}\right\rangle_\Psi-\left\langle\hat T_{zz}^{EM}\right\rangle_\Psi-\left(\left\langle\hat T_{00}^{EM}\right\rangle_\mathbf{C}-\left\langle\hat T_{zz}^{EM}\right\rangle_\mathbf{C}\right)\right]|f(t)|^2\ud t=\\
%\int_{-\infty}^\infty\left[\left\langle\hat T_{00}^\text{EM}\right\rangle_\Psi-\left\langle\hat T_{zz}^\text{EM}\right\rangle_\Psi-\left(\left\langle\hat T_{00}^\text{EM}\right\rangle_\mathbf{C}-\left\langle\hat T_{zz}^\text{EM}\right\rangle_\mathbf{C}\right)\right]f(t)\ud t
\mathcal D=\left\langle\hat T_{00}^\text{EM}\right\rangle_\Psi-\left\langle\hat T_{zz}^\text{EM}\right\rangle_\Psi-\left(\left\langle\hat T_{00}^\text{EM}\right\rangle_\mathbf{C}-\left\langle\hat T_{zz}^\text{EM}\right\rangle_\mathbf{C}\right)\geq -C(t_0),
\label{auxa2}
\end{equation}
\end{widetext}
%Or, rearranging,
%\begin{equation}
%\delta P\leq\delta\rho.
%    \label{col1}
%\end{equation}
because, for $f$ in Eq.~\eqref{Lorentzian}, $\int_{-\infty}^\infty f(t)\ \ud t=1$. Here,   %given by the right-hand side of inequality~\eqref{EMQEI} for the specified matrices $\mathbf P$ and $\mathbf M$.
\[
C(t_0)
=\!\!
\int_0^\infty  
\! \frac{\ud\alpha}{2\pi} 
\sum_\lambda \!
\int\ud^3\vec k \left|\widetilde{f^{\frac{1}{2}}}(\omega+\alpha)\right|^2 \!
\left[
\left|\mathcal E_z^{\lambda\vec k}\right|^2
\!\! +
\left|\mathcal B_z^{\lambda\vec k}\right|^2
\right]
\]
is a non-negative function of $t_0$.
To evaluate it, we explicitly compute $\widetilde{f^\frac{1}{2}}$ for the sampling function~\eqref{Lorentzian},
\begin{equation}
    \widetilde{f^\frac{1}{2}}(\omega)=2\sqrt{\frac{t_0}{\pi}}K_0(|\omega| t_0),
    \label{FTLorentzian}
\end{equation}
where $K_0$ is the modified Bessel function of the second kind. Hence,
\begin{eqnarray*}
    C(t_0)
    &=&2t_0\int_0^\infty  \ud\alpha \sum_\lambda\int\ud^3\vec k\ K_0^2[(\omega+\alpha)t_0]
    \\ 
    &\times&\left(\left|\mathcal E_z^{\lambda\vec k}\right|^2+\left|\mathcal B_z^{\lambda\vec k}\right|^2\right).
\end{eqnarray*}
Changing integration variables from $k_i$ to $\kappa_i=k_i/t_0$ and from $\alpha$ to $\varpi=\alpha/t_0$, and invoking Eq.~\eqref{ebcoulomb} for mode functions $\mathcal A^i_{\lambda\vec k}\propto \ue^{-\ui\omega t}/\sqrt{\omega}$, we obtain the scaling behavior for $C(t_0)$: 
$$
C(t_0)\propto t_0^{-4}.
$$ 
Thus, in the limit $t_0\to\infty$, $C(t_0)$ goes to zero and Eq.~\eqref{auxa2} reduces to the desired result~\eqref{col1}.

As a final remark, we can likewise obtain
\begin{equation}
%\delta\rho=
\left\langle\hat T_{00}^\text{EM}\right\rangle_\Psi-\left\langle\hat T_{00}^\text{EM}\right\rangle_\mathbf{C}\geq 0,
    \label{col2}
\end{equation}
which agrees with a similar result in two dimensions by Ford and Roman for the scalar field~\cite{Ford95} and can be readily interpreted by saying that $\mathbf C$ is the state of the lowest energy. For the proof, we write an expression similar to Eq.~\eqref{pres} for $\hat T_{00}^{EM}=\frac{1}{2}\sum_{i=1}^3(\hat E_i^2+\hat B_i^2)$. Similarly, the expression for 
$$
\displaystyle \int\left(\left\langle\hat T_{00}^{EM}\right\rangle_\Psi-\left\langle\hat T_{00}^{EM}\right\rangle_\mathbf{C}\right)f(t)\ud t
$$ 
has the same form as two copies the left-hand side of Eq.~\eqref{EMQEI} with $M_{ij}=\delta_{ij}$, one for $P^i=\mathcal E^i$ and the other for $P^i=\mathcal B^i$. Entirely analogous reasoning leads to Eq.~\eqref{col2}.
%Whereas for $Q=\frac{1}{\sqrt 2}\partial_\mu,\ \mu=0,1,2,3$ and $Q=\frac{1}{\sqrt 2}m$ leads to


\begin{thebibliography}{}
    
    \bibitem{Bondi57} 
    H. Bondi,
    Negative Mass in General Relativity,
    Rev. Mod. Phys. {\bf 29}, 423 (1957).
  
  \bibitem{Schoen79} 
    R. Schoen and S. Yau,
    On the proof of the positive mass conjecture in general relativity,
    Comm. Math. Phys. {\bf 65}, 45 (1979).
  
  \bibitem{Schoen81} 
    R. Schoen and S. Yau,
    Proof of the positive mass theorem. II,
    Comm. Math. Phys. {\bf 79}, 231  (1981).
  
  \bibitem{Witten81} 
    E. Witten,
    A new proof of the positive energy theorem,
    Comm. Math. Phys. {\bf 80}, 381 (1981).
    
  \bibitem{Archimedes14} 
  E. Calloni, M. De Laurentis, R. De Rosa, F. Garufi, L. Rosa, L. Di Fiore, G. Esposito, C. Rovelli, P. Ruggi, and F. Tafuri, 
    Towards weighing the condensation energy to ascertain the Archimedes force of vacuum, 
    Phys. Rev. D, {\bf 90}, 022002 (2014).
    
    \bibitem{Cerdonio15a} M. Cerdonio and C. Rovelli, A Casimir cannot cavity fly, arXiv:1406.1105 [gr-qc].

    \bibitem{Helfer98} 
    A. D. Helfer,
    {\textasciigrave}{O}perational{\textquotesingle} energy conditions,
    Class. Quantum Grav. {\bf 15}, 1169 (1998).
    
    \bibitem{Bekenstein13} 
    J. D. Bekenstein,
    If vacuum energy can be negative, why is mass always positive? Uses of the subdominant trace energy condition,
    Phys. Rev. D {\bf 88}, 125005 (2013).
    
 \bibitem{HawkingEllis}
    S. W. Hawking and G. F. R. Ellis, 
    {\em The Large Scale Structure of Space-Time} 
    (Cambridge Press, Cambridge, 1975).

\bibitem{Kuo93} 
    C.-I. Kuo and L. H. Ford,
    Semiclassical gravity theory and quantum fluctuations,
    Phys. Rev. D {\bf 47}, 4510 (1993).

\bibitem{Wald84}
    R. M. Wald, 
    {\em General Relativity} 
    (University of Chicago Press, Chicago, 1984).

\bibitem{Mostepanenko97}
    V. M. Mostepanenko, N. N. Trunov, and R. L. Znajek, 
    {\em The Casimir Effect and Its Applications} 
    (Clarendon Press, Oxford, 1997).

\bibitem{Milton04}
     K. A. Milton,
    The Casimir effect: recent controversies and progress,
    J. Phys. A  {\bf 37}, R209 (2004).

\bibitem{Birrell84}
     N. D. Birrell and P. C. W. Davies, 
    {\em Quantum Fields in Curved Space} 
    (Cambridge Press, Cambridge, 1984).

\bibitem{Milton08}
    K. A. Milton, S. A. Fulling, P. Parashar, A. Romeo, K. V. Shajesh, and J. A. Wagner,
    Gravitational and inertial mass of Casimir energy,
    J. Phys. A  {\bf 41}, 164052 (2008).
    
\bibitem{Brevik01}  
   I. Brevik and K. Pettersen,
    Casimir effect for a dielectric wedge,
    Ann. Phys. (N.Y.) {\bf 291}, 267 (2001).
    
\bibitem{MisnerThorneWheeler}
    C. W. Misner,  K. S.  Thorne,  and J. A. Wheeler, 
    {\em Gravitation} 
    (W. H. Freeman, San Francisco, 1973).

\bibitem{Flanagan91}
     E. Flanagan,
    Hoop conjecture for black-hole horizon formation,
    Phys. Rev. D {\bf 44}, 2409 (1991).
    
    \bibitem{Schwinger78}
    J. Schwinger, L. L. DeRaad, and K. A. Milton,
      Casimir Effect in Dielectrics,
     Ann. Phys {\bf 115}, 1 (1978).

    \bibitem{Bordag00} 
      M. Bordag, B. Geyer, G. L. Klimchitskaya, and V. M. Mostepanenko,
     Casimir force at both nonzero temperature and finite conductivity,
     Phys. Rev. Lett. {\bf 85}, 503 (2000).

\bibitem{Klimchitskaya09}
   G. L. Klimchitskaya, U. Mohideen, and V. M. Mostepanenko,
   The Casimir force between real materials: Experiment and theory,
    Rev. Mod. Phys.  {\bf 81}, 1827  (2009).
     
\bibitem{footnote} Alternatively, the force can also be computed by taking the derivative $\displaystyle -\frac{\partial}{\partial a}\int\ud^3x\ \left\langle \hat{T}^\text{EM}_{00}\right\rangle$.

\bibitem{Gies06}
H. Gies and K. Klingm\"uller, Casimir Edge Effects, Phys. Rev. Lett. {\bf 97}, 220405 (2006).

\bibitem{Brown69} 
     L. S. Brown and G. J. Maclay,
     Vacuum Stress between Conducting Plates: An Image Solution,
    Phys. Rev. {\bf 184}, 1272 (1969).
    
\bibitem{deWitt75}
    B. S. DeWitt,
    Quantum field theory in curved spacetime,
    Phys. Rep. {\bf 19}, 295 (1975).
    
\bibitem{Hagen01}
    C. Hagen, 
    Cutoff dependence of the Casimir effect, 
    Eur. Phys. J. C \textbf{19}, 677 (2001).
     
%\bibitem{Cerdonio15}
%    M. Cerdonio and C. Rovelli,
%    Casimir effects are not an experimental demonstration that free vacuum gravitates: connections to the Cosmological Constant Problem,
%    Int. J. Mod. Phys. D {\bf 24}, 1544020 (2015).

\bibitem{Morris88}
    M. S. Morris and K. S. Thorne,
    Wormholes in spacetime and their use for interstellar travel: A tool for teaching general relativity,
    Am. J. Phys. {\bf 56}, 395 (1988).

\bibitem{Pfenning01}
      M. J. Pfenning,
     Quantum inequalities for the electromagnetic field,
     Phys. Rev. D {\bf 65}, 024009 (2001).

\bibitem{Ford95}
    L. H. Ford and T. A. Roman,
    Averaged energy conditions and quantum inequalities,
    Phys. Rev. D {\bf 51}, 4277 (1995).

%\bibitem{Saharian04}
%\bruno{A. A. Saharian, Energy-momentum tensor for a scalar field on manifolds with boundaries, Phys. Rev. D {\bf 69}, 085005 (2004).}

\end{thebibliography}
\end{document}